\documentclass[%
 reprint,
superscriptaddress,
 aps,
 prl
]{revtex4-2}

\usepackage{xfrac}
\usepackage{braket}
\usepackage{physics}
\usepackage{graphicx}% Include figure files
\usepackage{dcolumn}% Align table columns on decimal point
\usepackage{bm}
\usepackage{subcaption}
\usepackage{graphicx}
\DeclareMathSizes{10}{10}{4}{4}
\usepackage[colorlinks=true, allcolors=blue]{hyperref}
\usepackage{floatrow}
\newcommand{\labelphantom}[1]{%
  \parbox{0pt}{\phantomsubcaption\label{#1}}%
}
\newcommand{\eqn}[1]{Eq.~(\ref{#1})}
\newcommand{\Eqn}[1]{Equation~(\ref{#1})}

\begin{document}
\preprint{APS/123-QED}

\title{On the origin of exponential operator growth in Hilbert space}% Force line breaks with \\

\author{Vijay Ganesh Sadhasivam}\email{vgs23@cam.ac.uk}
\affiliation{Max-Planck-Institut für Physik komplexer Systeme, Nöthnitzer Straße 38, 01187 Dresden, Germany}
\affiliation{Yusuf Hamied Department of Chemistry, University of Cambridge, Lensfield Road, Cambridge, CB2 1EW, UK}

\author{Jan M. Rost}\email{rost@pks.mpg.de}
\affiliation{Max-Planck-Institut für Physik komplexer Systeme, Nöthnitzer Straße 38, 01187 Dresden, Germany}

\author{Stuart C.\ Althorpe}\email{sca10@cam.ac.uk}
\affiliation{Yusuf Hamied Department of Chemistry, University of Cambridge, Lensfield Road, Cambridge, CB2 1EW, UK}

% Work to do
% 1. References
% 2. Figures
% 3. Formula for random matrix

\date{\today}% It is always \today, today,

\begin{abstract}
\textcolor{black}{The question of thermalization in quantum many-body systems has long been studied through the properties of matrix elements of operators corresponding to local observables. More recently, the focus has shifted to the dynamics of operators, which lead to seminal works proposing universal bounds on the rate of operator growth. In this work, we unify these two approaches: we
show that exponential operator growth in Hilbert space, as measured by Krylov complexity, is governed by an exponential off-diagonal decay of the operator matrix elements in the system eigenbasis. When this decay is algebraic or slower, the growth rate saturates the universal bound, thereby establishing a  microscopic origin of operator growth which is independent of chaos, dimensionality or the presence of many-body interactions. }
\end{abstract}

\maketitle

\textcolor{black}{The question of if and how quantum many-body systems approach thermal equilibrium, has been a longstanding and central theme of both theoretical \cite{polkovnikov2011colloquium, eisert2015quantum} and experimental research \cite{trotzky2012probing, kaufman2016quantum}. Theoretical studies \cite{rigol2008thermalization, d2016quantum, deutsch2018eigenstate, srednicki1994chaos} spanning the last three decades have related the equilibrium properties of systems and the fluctuations around them to matrix elements of operators corresponding to local observables \cite{beugeling2015off, leblond2019entanglement}. }   

\textcolor{black}{In recent years, more operator-centric descriptions of the thermalisation process have emerged, with a renewed focus on the irreversible spread of quantum correlations in a system with time referred to as quantum `scrambling' \cite{von2018operator, nahum2018operator, xu2019locality, shenker2014black}. These studies consider operator dynamics explicitly using time-correlation functions, providing dynamical quantifiers for operator growth \cite{maldacena2016bound, swingle2018unscrambling,parker2019universal, nandy2025quantum}. Most recent work \cite{parker2019universal, camargo2024spectral, huh2024spread} has been concerned with operator growth in \textit{Hilbert space} where an exponential operator growth, say at the rate $2\alpha$, is bounded by a universal (i.e. system-independent) bound
\begin{align}\label{alpha-bound}
    \alpha(T) \leq {\pi k_BT}/{\hbar}
\end{align}
at finite temperatures. This bound, similar to a related bound on chaos \cite{maldacena2016bound}, constrains the speed at which information spreads in quantum systems.}

\textcolor{black}{The purpose of this letter is to analyse the conditions for exponential
growth of operators in their respective Hilbert space under thermal equilibrium and the maximality of this growth rate. Specifically, we establish that exponential operator growth is caused solely by the energy dependence of the off-diagonal elements of the operator in the system eigenbasis. This diagnostic provides a unifying framework that consolidates the two distinct approaches to thermalisation and further clarifies the lack of a simple correspondence between ‘fast scrambling’ or exponential operator growth and quantum chaos \cite{xu2020does, bhattacharjee2022krylov, camargo2024spectral, kidd2020thermalization, sadhasivam2023instantons, michel2024quasiclassical, hashimoto2020exponential}. Moreover, it predicts examples of many-body systems where operators do not grow exponentially, and single particle systems where operators grow exponentially, possibly at the maximal rate.}

The dynamics of an operator $\hat{O}$ is given by its Heisenberg equation of motion:
\begin{align}
    \frac{\partial}{\partial t}\hat{O}(t) = \frac{i}{\hbar} [\hat{H},\hat{O}(t)] := i\hat{\mathcal{L}} \hat{O}(t)
\end{align}
where $\hat{\mathcal{L}} = [\hat{H},\cdot]/\hbar$ is the quantum Liouvillian superoperator. Since the equation is linear, the formal solution to this equation $\hat{O}(t) = e^{i\hat{\mathcal{L}}t} \hat{O}(0)$ can be completely described in the Hilbert space spanned by the basis $\mathcal{S}_{O} = \{ \mathcal{L}^n \hat{O}\}_{n=0}^{\infty} $ of nested commutators. At finite temperature $\beta=1/k_BT$, the (Wightman) auto-correlation function
\begin{align}\label{TCF}
     C(t) = \frac{1}{Z} \text{Tr} [e^{-\beta \hat{H}/2}\hat{O}(0) e^{-\beta \hat{H}/2}\hat{O}(t)],
\end{align}
which quantifies the dynamical correlation of $\hat{O}$ defines the inner product $\langle \hat{O} | \hat{O}(t)\rangle_{\beta}$.

\begin{figure}
\labelphantom{fig1a}
\labelphantom{fig1b}%
\labelphantom{fig1c}%
\labelphantom{fig1d}
\labelphantom{fig1e}%
\labelphantom{fig1f}%
    \centering
    \includegraphics[width=\linewidth]{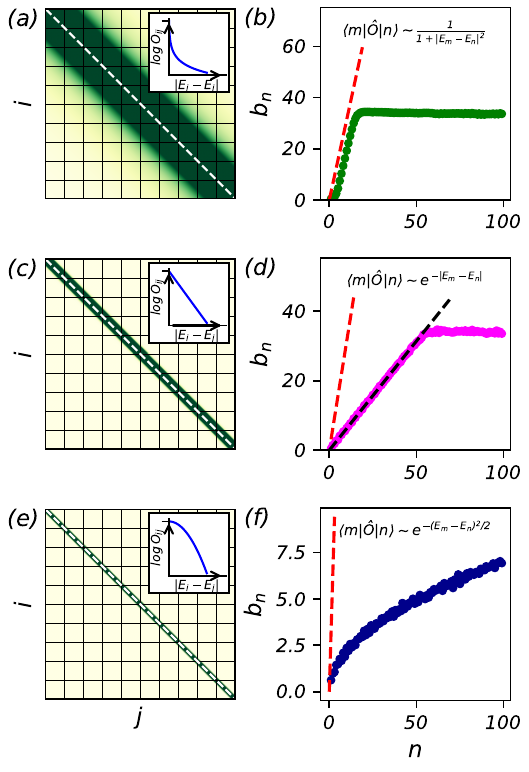}
    \caption{Illustration of (a) power-law, (c) exponential and (e) Gaussian decay of off-diagonal matrix elements of an operator in the energy eigenbasis of a random-matrix ensemble. (b), (d) and (f) The corresponding Lanczos sequence averaged over the ensemble \cite{tang2023operator}. The red dashed line indicates the maximal growth as in \eqref{alpha-bound}. }
\end{figure}

Following \cite{parker2019universal}, we orthonormalise the basis $\mathcal{S}_{O}$ using the \textit{Lanczos algorithm} (which follows a Gram-Schmidt procedure) to yield the \textit{Krylov} basis $\mathcal{K}_O = \{\hat{O}_n\}_{n=0}^{\infty}$ along with a sequence of Lanczos coefficients $\{b_n\}_{n=0}^{\infty}$ which embodies all the information concerning the increasing complexity of the time-evolved operator $\hat{O}(t)$. The operator $\hat{\mathcal{L}}$ is tridiagonal in this basis and hence the time-dependent occupancy
$\varphi_n(t) = i^{-n} \langle \hat{O}_n | \hat{O}(t) \rangle$ in each of the `site' $n$ in the Krylov basis is given by the following discrete time-dependent Schrodinger equation (DTDSE):
\begin{align}
    \frac{\partial \varphi_n(t)}{\partial t} = -b_{n+1}\:  \varphi_{n+1}(t) + b_n \: \varphi_{n-1}(t) 
\end{align}
with $\varphi_n(0) = \delta_{n0} $. The \textit{Krylov complexity} measures the spread of the operator in this basis and is given as a weighted average of the occupancy at each `site' of the lattice formed by $\hat{O}_n$:
\begin{align}\label{K-comp}
     C_{O}^K(t) = \sum_{n=0}^{\infty} n|\varphi_n(t)|^2,
\end{align}
When
\begin{align}\label{cot}
C_{O}^K(t) \sim e^{2\alpha t}
\end{align}
the operator is considered to grow exponentially in
Hilbert space. \Eqn{cot} can be shown to be equivalent to linear growth
\begin{align}
    b_n\sim\alpha n
    \end{align}
of the Lanczos coefficients $b_n$, and to exponential growth of the form
\begin{align}\label{mu_max}
    \mu_{2n} \sim \left ( \frac{4n\alpha}{e \pi}  \right)^{2n} 
\end{align}
of the moments
\begin{align}\label{moments_def}
     \mu_{2n}  = {\partial^{2n}\over\partial t^{2n}} C(t)
\end{align}
of the time-correlation function $C(t)$. When the operator growth is exponential, the universal operator growth hypothesis \cite{parker2019universal} predicts that the growth rate is bounded as in \eqref{alpha-bound} at finite temperature.

In what follows, we will represent $\hat{O}$ in the basis set of eigenstates of $\hat H$. The off-diagonality of $\hat{O}$ in this basis is obviously crucial to its growth, since a diagonal operator is static (and hence its Lanczos sequence vanishes entirely). For a harmonic oscillator, the position operator $\hat x$ has matrix elements 
\begin{align}
    x_{lk} = \sqrt{\frac{\hbar}{2m\omega}} (\sqrt{k+1}\:\delta_{l,k+1} + \sqrt{k} \: \delta_{l,k-1})
\end{align}
between energy eigenstates $\ket{l}$ and $\ket{k}$, which cause the Lanczos sequence to terminate at $n=2$. But in general, the (symmetry-allowed) off-diagonal matrix elements of $\hat{O}$ are non-zero and typically decay as some function of the off-diagonal energy-difference $\omega$.

\textcolor{black}{For instance, the matrix elements of a generic, interacting many-body system is usually well-described by the \textit{Eigenstate Thermalisation Hypothesis}} (ETH) \cite{srednicki1994chaos,deutsch2018eigenstate}:
\begin{align}\label{ETH-gen}
    O_{lk} := \langle l | \hat{O} | k \rangle =
O(\bar{E}) \, \delta_{lk}
+ e^{-S(\bar{E})/2} \, f_O(\bar{E}, \omega) \, R_{lk}
\end{align}
where $\{\ket{E_n}\}_{n=0}^{\infty}$ corresponds to the many-body eigenbasis, $\bar{E}=(E_l+E_k)/2$ and $\omega=E_l-E_k$, $S(\bar{E})$ is the entropy of the system at energy $\bar{E}$, $R_{lk} \sim \mathcal{N} (0, 1)$,  and $f_O(\bar{E},\omega )$ and $O(\bar{E})$ are smooth functions of $\bar{E}$ and $\omega$. Additionally, if the system is `quantum chaotic' in the sense that its energy-level spacing follows Wigner-Dyson level statistics and the operator $\hat{O}$ is described by a random matrix, then we have $O(\bar{E}) = 0$ and $f_O(\bar{E}, \omega ) = 1$ \cite{d2016quantum}, with a constant entropy function $S(E)=\ln{\mathcal{D}}$ (where $\mathcal{D}$ is the Hilbert-space dimension of $\hat O$); i.e. 
\begin{align}\label{RMT-mat}
    O_{lk} = \frac{1}{\sqrt{\mathcal{D}}} R_{lk}
\end{align}
Expanding the moments  $\mu_{2n}$ of \eqn{mu_max} in the basis of energy eigenstates, we obtain
\begin{align}\label{moments}
    \mu_{2n} = \frac{1}{Z}\sum_{l=1}^\infty\sum_{k=1}^{\infty} (E_l-E_k)^{2n} e^{-\beta (E_l+E_k)/2} |O_{lk}|^2
\end{align}
We can rewrite this expression as 
\begin{align}
     \mu_{2n} &\approx \frac{\int_0^\infty d\bar{E} \: e^{S(\bar{E})-\beta \bar{E}}\int_{-2\bar{E}}^{2\bar{E}}d\omega\: \omega^{2n} |O(\bar{E},\omega)|^2}{\int_0^{\infty} d\bar{E}\: e^{S(\bar{E})-\beta \bar{E}}}
\end{align}
provided the number of thermally accessible states is sufficiently high that we can replace the sum by an integral using \cite{d2016quantum} (also, see SI)
\begin{align}\label{sumint}
    \sum_{l=1}^\infty \rightarrow \int dE_l \: e^{S(E_l)},
\end{align}

\begin{figure*}[!htb]
\labelphantom{fig2a}%
    \labelphantom{fig2b}%
    \labelphantom{fig2c}
    \centering
    \includegraphics[width=\linewidth]{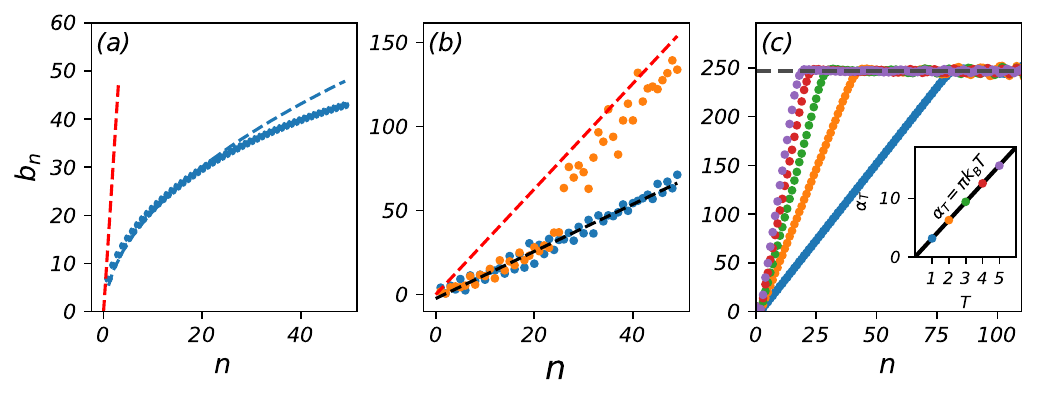}
    \caption{Plots of Lanczos sequence for (a) $\hat{O} =\hat{x}^{100}$ in a harmonic oscillator $V(x)=x^2/2$ (blue dots); the dashed line corresponds to a fictional operator with approximate matrix elements as in \eqref{xk-gauss-approx} indicating Gaussian decay of matrix elements. (b) $\hat{O}=\hat{x}$ in a quartic anharmonic oscillator $V(x)=x^4$ (yellow) - the artificial transition in the growth rate at $n\sim 25$ is due to the limit in numerical precision while  computing position matrix elements; blue dots correspond to a fictional operator with (sub)exponentially decaying matrix elements as in \eqref{landau-mat}. (c) $\hat{O}=\hat{x}$ in a one-dimensional infinite box (length $L=10$) at various temperatures. Inset: Plot of the growth rate against temperature indicating maximal growth.   }
    \label{fig:placeholder}
\end{figure*} 

Averaging over the random variable $R_{lk}$, we obtain 
\begin{align}
    \langle \mu_{2n} \rangle &\approx 2\int_{0}^{\infty}d\omega \: \omega^{2n} e^{-\beta|\omega|/2} \nonumber\\
    &\sim \left( \frac{4n}{\beta e}\right)^{2n}\label{maxxy}
\end{align}
which corresponds to exponential growth of $C_{O}^K(t)$ at the maximal rate $\alpha=\pi/\beta$ permitted by \eqn{alpha-bound} (see \eqn{mu_max}) setting $\hbar=1$.
Numerical propagation of the Lanczos algorithm confirms this prediction (indicating the validity of \eqn{sumint}), with $b_n$ growing linearly at the maximal growth rate---see Fig.~\ref{fig1a}.

If we repeat the steps that led to \eqn{maxxy}, but with the structure function $f_O$ exponentially suppressed in $\omega$ as: 
\begin{align}
    f_O(\bar{E},\omega) \sim e^{-\gamma |\omega|}
\end{align}
we obtain
\begin{align}
    \mu_{2n} &\sim  \left[ \frac{4n}{(\beta+4\gamma)e}\right]^{2n}
\end{align}
i.e.\ $\hat O$ is still predicted to grow exponentially, but at a sub-maximal rate $\alpha=\pi/(\beta+4\gamma)$, which is again confirmed by the numerical results in Fig.~\ref{fig1b}. \textcolor{black}{Such exponential (or Gaussian) suppression is a result of higher-order many-body processes required for transitions between states with widely separated energies \cite{d2016quantum,leblond2019entanglement,beugeling2015off,khatami2013fluctuation}.}

If $f_O$ decays faster than exponentially as a function of $\omega$, $\mu_{2n}$ would grow slower than $\sim n^{2n}$, which would correspond to slower than exponential growth in $C_{O}^K(t)$ (and sub-linear growth with in $b_n$). Thus, the interplay between quantum thermal fluctuations and the internal fluctuations of operators \cite{d2016quantum} decides the operator growth rate, thereby indicating a \textit{purely statistical} origin of the bound on the growth rate, similar to that on the quantum Lyapunov exponent \cite{sadhasivam2023instantons,pappalardi2022quantum,murthy2019bounds,sadhasivam2024thermal}. Numerical propagation of the Lanczos algorithm also confirms this prediction as shown in Fig.~\ref{fig1c}.

These random-matrix results suggest a simple diagnostic: the growth rate of $\hat{O}$ is determined by its off-diagonal matrix elements. Exponential or slower decay in $\omega$ yields exponential growth, with algebraic decay giving the maximal rate \eqn{alpha-bound}; otherwise the growth will be slower than exponential. We demonstrate that this mechanism is not restricted to a many-body/random-matrix context by considering single- and few-body examples which show that this simple picture appears to hold in more realistic models.

First, consider the growth of the operator $\hat{\mu} = \hat{x}^q$ for a harmonic oscillator. The off-diagonal matrix elements of $\hat{\mu}$ can be shown to decay as
\begin{align}\label{xk-gauss-approx}
    \langle l |\hat{\mu}|k\rangle \sim  \exp\left (-\frac{|l-k|^2}{2q}\right),
\end{align}
(see the SI). On the basis of the above, we therefore expect the operator growth to be slower than exponential, and this is confirmed numerically in Fig.~\ref{fig2a} (using $m=1$ and $\beta=1$). Let us now define an operator $\hat{O}$ such that the matrix elements (in the harmonic-oscillator eigenbasis) take the form  $\hat{O}_{lk}=R_{lk}$. Averaging over the randomness of $R_{lk}$, we obtain
\begin{align}
     \langle\mu_{2n}\rangle &= 2(\hbar\omega)^{2n} \text{Li}_{-2n}(e^{-\beta\hbar\omega/2})
\end{align}
where $\text{Li}_{-n}(z)$ is the negative polylogarithm function. As $\beta\hbar\omega \rightarrow0 $, we obtain
\begin{align}
    \langle\mu_{2n}\rangle \rightarrow \frac{\omega^{2n}}{(\beta\omega/2)^{2n+1}}(2n)! \rightarrow \left ( \frac{4n}{\beta e}\right)^{2n}
\end{align}
In other words, even a harmonic oscillator can give rise to exponential operator growth that maximises \eqn{alpha-bound} (in the high-temperature limit), provided the operator $\hat{O}$ decays sufficiently slowly as a function of $\omega$.

\begin{figure}
    \labelphantom{fig3a}
    \labelphantom{fig3b}
    \centering
    \includegraphics[width=\linewidth]{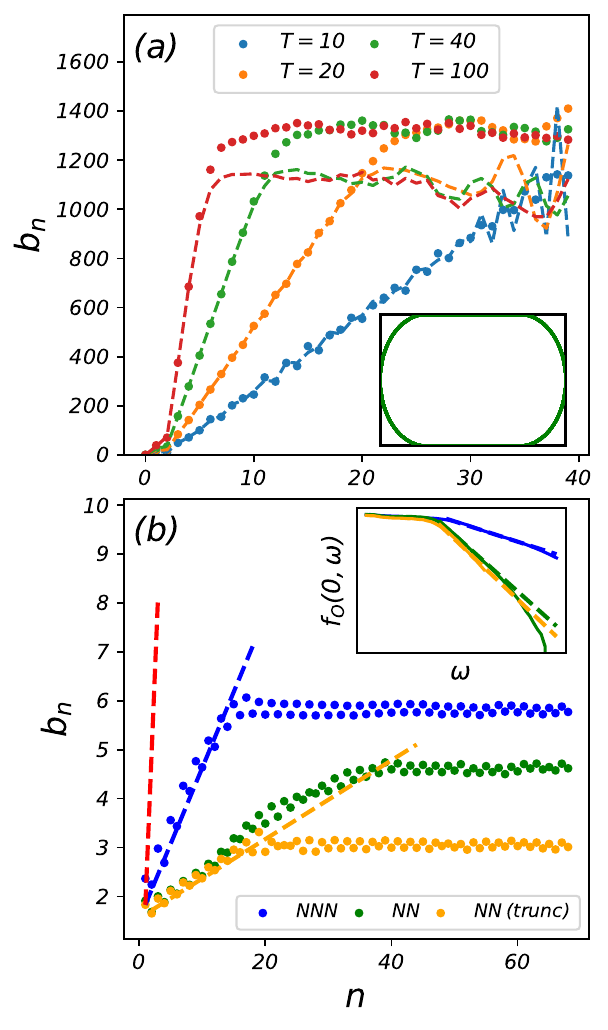}
    \caption{(a) Lanczos sequence for $\hat{O}=\hat{x}$ in a stadium billiard (dots) with area $1$ and the parameter length-radius ratio $a/R=1.0$ as in \cite{camargo2024spectral} at various temperatures. The dashed lines correspond to the Lanczos sequence for a rectangular box of similar dimensions as shown in the inset. (b) Lanczos sequence for the flip-flop operator $\hat{B}$ in \eqref{B-flip-flop} for the XXZ model with NN (green) and NNN (blue) coupling. The yellow dots corresponds to the operator truncated to an eigenbasis of size $N=2000$. Inset: Plots of the structure function at $E=0$ for each of the cases, showing exponential (NNN) and Gaussian (NN) decay.  }  
\end{figure}

One could argue that this last example is artificial, but more realistic examples of exponential or even maximal growth in non-chaotic systems can easily be found. For example, consider a one-dimensional system with potential 
\begin{align}\label{Vx}
    V(x) = x^{p},
\end{align}
where $p>2$ is an even integer. Using WKB theory \cite{landau2013quantum,cornwall1993semiclassical}, the energy eigenvalues can be approximated by
\begin{align}
    E_k \sim k^{2p/(p+2)}
\end{align}
and the matrix elements of $\hat x$ by
\begin{align}\label{landau-mat}
    x_{lk} \approx \exp ({-L_{lk}}) \quad \quad |l-k|>>1
\end{align}
where
\begin{align}
    L_{lk} \sim |E_l-E_k|^{(p+2)/2p},
\end{align}
for large $|n-m|$ (see the SI) and therefore $x_{lk}$ decays sub-exponentially with $\omega$ (since $(p+2)/2p\leq 1$).
On the basis of the above, we therefore expect $O=\hat{x}$ to grow exponentially but not maximally. This prediction is confirmed numerically in Fig.~\ref{fig2b}, which was generated by propagating the Lanczos algorithm (with $m=1$, $\beta=1$) with the $x_{lk}$ evaluated (i) using the numerically calculated eigenfunctions  and (ii) analytically, using the semiclassical approximation of \eqn{landau-mat}. Note the presence of a  potentially  misleading numerical artefact, in which the growth rate obtained in (i) appears to jump upwards at about $n=25$ to the maximal rate. This jump happens because the off-diagonal matrix elements for $\omega>75$ are dominated by machine errors and thus resemble random matrix elements (which do not decay with $\omega$).

As $p\rightarrow\infty$, the system of \eqn{Vx} tends to a particle in a one-dimensional box, for which 
\begin{align}
    x_{mn} = \begin{cases} \frac{8nm}{\pi^2(n^2-m^2)^2} \;\;\; &m+n \text{ odd} \\
    0 &\text{otherwise}  
    \end{cases}
\end{align}
Since these elements decay  algebraically with $\omega$, we expect that $\hat x$ will grow at the maximum rate $\alpha=\pi/\beta$. Numerical propagation of the Lanczos algorithm over a range of temperatures confirms this prediction over a wide temperature range---see Fig.~\ref{fig2c}. 

Since a particle in a one-dimensional box gives maximal exponential growth, it  immediately follows that a particle in a two-dimensional rectangular box, being a separable system, also gives maximal growth. The growth of $\hat{x}$ for a particle in a rectangular box is plotted in Fig.~\ref{fig3a}, where it is shown to closely follow the growth reported in ref.~\cite{camargo2024spectral} for a stadium billiard (until the growth stops at saturation). This shows that the maximal growth achieved by the latter results from confinement rather than quantum chaos, which leads to an algebraic decay of $x_{lk}$ with $\omega$.

\textcolor{black}{To test our theory in a truly many-body setting, we now consider the XXZ chain, with periodic boundary conditions, a canonical many-body model which can be tuned between integrable and chaotic regimes}. We consider the growth of the next-nearest-neighbour flip-flop operator
\begin{align}\label{B-flip-flop}
    \hat{B} = \frac{1}{L}\sum_{i=1}^L \hat{S}_i^+ \hat{S}_{i+2}^{-} + \hat{S}_i^-\hat{S}_{i+2}^+
\end{align}
for an XXZ Hamiltonian with nearest-neighbour (NN) and next-nearest-neighbour (NNN) coupling
\begin{equation}
\begin{split}
\hat{H}_{\text{XXZ}} = J_1 \left(\sum_{i=1}^L  \hat{S}_i^x \hat{S}_{i+1}^x + \hat{S}_i^y \hat{S}_{i+1}^y + \Delta_1 \hat{S}_{i}^z\hat{S}_{i+1}^z \right) +\\ J_2 \left(\sum_{i=1}^L  \hat{S}_i^x \hat{S}_{i+2}^x + \hat{S}_i^y \hat{S}_{i+2}^y + \Delta_2 \hat{S}_{i}^z\hat{S}_{i+2}^z \right)
\end{split}
\end{equation}
where $i$ satisfies periodic boundary conditions. The parameter $J_1$ controls the nearest-neighbour (NN) coupling between sites, and $J_2$ controls the next-nearest-neighbour (NNN) coupling. We consider two different parameterisations of this model, an `NN' model, in which $L=20$, $J_1=1.0$, $\Delta_1=0.55, \Delta_2=0.5$, and $J_2=0$, and an `NNN' model which uses the same parameters except that $J_2=1$. The NN model is integrable and the matrix elements of $\hat B$ decay exponentially with $\omega$ for $N<2000$ and have Gaussian decay for $N>2000$; the NNN model is chaotic \cite{kudo2005level,leblond2019entanglement} (with Wigner-Dyson statistics) and the matrix elements of $\hat B$ decay exponentially. We thus expect the NN model to give exponential growth initially, followed by slower growth, and the NNN model to give exponential growth.
These predictions are confirmed by the numerical results in Fig.~\ref{fig3b}, generated by propagating the Lanczos algorithm at $\beta=1$. To check that it is the $\omega$-dependence of $\hat B$ which produces the exponential growth in the NNN model (not the chaoticity), we rerun the NN calculations with the $\hat B$ matrix artificially truncated at $N=2000$ (so that all non-zero elements decay exponentially with $\omega$); as expected, the growth is purely exponential (Fig.~\ref{fig3b}). The exact diagonalisation of the many-body Hamiltonian was performed using the QuSpin package \cite{weinberg2017quspin}.

\textcolor{black}{In summary, our work identifies the $\omega$-dependence of the operator matrix elements as the diagnostic criteria for its exponential growth in Hilbert space. This dependence in $\omega$ measures the strength of coupling between a pair of eigenstates through the operator. Although many-body quantum chaos usually leads to an exponential or slower decay in $\omega$ necessary for exponential/maximal operator growth, one can construct models for which the decay is faster \cite{leblond2019entanglement, jansen2019eigenstate} and operator growth is slower than exponential. The strong coupling necessary for exponential operator growth can also result from single-body features such as infinite confinement and tunneling, as shown in this letter.}

\textcolor{black}{Our findings obviate the need for an explicit propagation of the Lanczos algorithm to predict and understand exponential (and maximal) operator growth in many-body quantum systems. By connecting it to the decay of matrix elements, our study also provide a direction towards the experimental measurement of operator growth \cite{mallayya2019heating}. An understanding of operator growth is central to understanding concepts such as eigenstate thermalisation and scrambling in many-body systems and this work relates it to properties of single-body systems, a connection which has been long sought after \cite{richter2022semiclassical,liao2020many}. The theory presented in this letter also motivates the design of quantum quenches in anharmonic one-dimensional continuous quantum systems at finite temperatures to probe the nature of the resultant nonequilibrium state \cite{will2010time,trotzky2012probing}.}

VGS and SCA acknowledge funding from a Leverhulme Trust research grant. VGS acknowledges funding from the Max Planck Society. VGS, JMR and SCA contributed equally to this work. The authors also acknowledge several helpful discussions with Andrea Pizzi.

\bibliographystyle{apsrev4-2}
\bibliography{citations}

\end{document}

\maketitle

\section{Gaussian decay of matrix elements in harmonic oscillator}

We approximate the matrix elements of the operator $\hat{O}=\hat{x}^q$ considered in the main text. Since the matrix elements of the position operator span just the sub- and super-diagonal in the energy eigenbasis with
\begin{align}
    x_{lk} = \sqrt{\frac{\hbar}{2m\omega}} (\sqrt{k+1}\:\delta_{l,k+1} + \sqrt{k} \: \delta_{l,k-1}),
\end{align}
we can approximate it with an operator $\hat{u}$ with the following matrix elements:
\begin{align}
    u_{lk} = \delta_{l,k+1} + \: \delta_{l,k-1}
\end{align}
In the calculation of moments $\mu_{2n}$ which increase steeply with $n$, we can hence approximate $\hat{x}^q$ with $\hat{u}^q$, whose matrix elements are:

    \begin{align}\label{O2Kmatrixelts}
    [\hat{u}^q]_{lk} = \begin{cases}
    0 \;\;\; \text{if $|l-k| > q$ }
    \\
        \begin{pmatrix}
        q\\(q-|l-k|)/2 
    \end{pmatrix} \:\:\: \text{if} \:q-|l-k|\: \text{is even and } l+k\geq q\\
    \begin{pmatrix}
        q\\(q-|l-k|)/2 
    \end{pmatrix} - 
    \begin{pmatrix}
        q\\(q-(l+k))/2 -1
    \end{pmatrix}  \:\:\:\text{if} \:q-|l-k|\: \text{is even and } l+k< q \\
    0 \;\;\; \text{otherwise}
    \end{cases}
\end{align}
Ignoring the boundary term, we have:
\begin{align}
     \begin{pmatrix}
        q\\(q-|l-k|)/2 
    \end{pmatrix} \approx 2^q \sqrt{\frac{2}{\pi q}} \: e^{-|l-k|^2/(2q)}
\end{align}
from Stirling approximation, for $q>>1$ and $|l-k|<<q$. Thus, we have Gaussian decay of matrix elements for $\hat{O}=\hat{x}^q$:
\begin{align}
    x_{lk} \sim e^{-|l-k|^2/(2q)}
\end{align}

\section{Semiclassical matrix elements for anharmonic oscillator}

Landau \cite{landau2013quantum}  and Voloshin \cite{voloshin1991strong} have provided a semiclassical estimate \cite{cornwall1993semiclassical} for the matrix elements $\bra{n}\hat{x}\ket{m}$ of bounded potentials when $|n-m|>>1$:
\begin{align}
    \bra{n}\hat{x}\ket{m} \sim e^{-L_{nm}}
\end{align}
where 
\begin{align}
    L_{nm} = \int_{E_m}^{E_n} \int_{a(u)}^{\infty} \frac{du}{\sqrt{2(V(x) - u))}}\:dx
\end{align}
where $a(u)$ is the classical outer turning point at the energy $u$. For anharmonic potentials $V(x) = x^p$, we get:
\begin{align}
    a(u) = u^{1/p}
\end{align}
which gives:
\begin{align}
    L_{nm} = \int_{E_m}^{E_n} \int_{u^{1/p}}^{\infty} \frac{du}{\sqrt{2(x^p - u))}}\:dx
\end{align}
Consider
\begin{align}
    I(u) = \int_{u^{1/p}}^{\infty} \frac{dx}{\sqrt{2(x^p - u))}}
\end{align}
Set $y = x^p -u$, so that
\begin{align}
    dy &= px^{p-1} dx \Rightarrow dx = \frac{1}{p}x^{1-p} dy\\
    x &= (y+u)^{1/p}  \\
    dx &=\frac{1}{p} (y+u)^{(1-p)/p}dy
\end{align}
 and 
\begin{align}
    I(u) &= \int_{0}^{\infty} \frac{dy}{p\sqrt{2y} (y+u)^{(p-1)/p}} \\
    &=\frac{\sqrt{2}}{p} \int_0^{\infty}  \frac{dv}{(v^2+u)^{(p-1)/p}}\\
    &= \frac{\sqrt{2}}{p} u^{1/2-(p-1)/p}\int_0^{\infty}  \frac{dv}{(v^2+1)^{(p-1)/p}}
\end{align}
Thus, we get:
\begin{align}
    I(u) \sim u^{(2-p)/(2p)}
\end{align}
In fact, it is known that:
\begin{align}
    \int_0^{\infty}  \frac{dv}{(v^2+1)^a} = \frac{\sqrt{\pi}}{2} \frac{\Gamma(a-1/2)}{\Gamma(a)}
\end{align}
so we can evaluate this exactly. Now,
\begin{align}
    L_{nm} &= \int_{E_m}^{E_n} I(u) du\\
    &= \left. u^{(p+2)/(2p)} \right|_{E_m}^{E_n}
\end{align}

The energy levels for an oscillator $V(x) = x^p$ can also be computed using the Bohr-Sommerfeld quantisation condition:
\begin{align}
    \oint p \:dq = 2\pi \hbar\left( n+ \frac{1}{2} \right)
\end{align}
For an anharmonic oscillator $V(x)=x^p$, the turning points at energy $E$ are $x_E = \pm E^{1/p}$ and the momentum as a function of energy and position is given as:
\begin{align}
    p(x) = \sqrt{2m(E-x^p)}
\end{align}
thereby yielding:
\begin{align}
   \oint p \:dq &= 2 \int_{-E^{1/p}}^{E^{1/p}} dx \: \sqrt{2m(E-x^p)} = 4 E^{1/p} \sqrt{2mE}\int_0^1 dy\:\sqrt{1-y^p} \\
   &=4 E^{(2+p)/2p} \sqrt{2m} \:G(p)
\end{align}
where
\begin{align}
    G(p) =  \frac{\Gamma(1+\frac{1}{p}) \Gamma(\frac{3}{2})}{\Gamma(\frac{1}{p}+\frac{3}{2})}
\end{align}
Thus we have:
\begin{align}
    E_n &= \left(\frac{\pi \hbar}{2\sqrt{2m}\:G(p)}\right)^{2p/(p+2)} \left(n+\frac{1}{2}\right)^{2p/(p+2)}\\
    &\sim C(p) \: n^{2p/(p+2)}
\end{align}
which gives:
\begin{align}
    L_{nm} \approx J(p) |n-m| \sim |E_n^{(p+2)/2p} - E_m^{(p+2)/2p}|
\end{align}
where 
\begin{align}
J(p) = \frac{\Gamma((p-2)/(2p))}{\Gamma((p-1)/p)} \frac{\sqrt{2\pi}}{p+2}\:C(p)
\end{align}
It is straightforward to see that $J(p)\rightarrow 0$ as $p\rightarrow\infty$, i.e. in the limit of particle in an infinite box. That is not surprising because the position matrix elements do not decay exponentially in a one-dimensional box.

\section{Approximation of the operator growth rate}
The operator growth rate can be found from the dependence of the moments $\mu_{2n}$ on $n$. We know that
\begin{align}
    \mu_{2n} = \frac{1}{Z}\sum_{m=0}^{\infty}\sum_{l=0}^{\infty} (E_l -E_m)^{2n} e^{-\beta(E_m+E_l)/2}  |\bra{l}\hat{O}\ket{m}|^2
\end{align}
We can approximate this sum by an integral by the following identification \cite{d2016quantum, Claeys2023_ETH}:
\begin{align}
    \sum_n \rightarrow \int dE_n  \:e^{S(E_n)}
\end{align}
where $S(E)$ is the entropy at energy $E$ (i.e. we are including a factor to account for density of states at energy $E$) and get:
\begin{align}
    \mu_{2n} \sim \int_0^{\infty} dE_l \int_0^{\infty} dE_m \: (E_l -E_m)^{2n} e^{-\beta (E_m+E_l)/2 + S(E_l)+S(E_m)} |\hat{O}_{lm}|^2
\end{align}
We can now change to \textit{Wigner variables} $\bar{E}=(E_l+E_m)/2$ and $\omega=E_l-E_m$, so that $E_{l,m} = E\pm \omega/2$ and 
\begin{align}
    \mu_{2n} \sim  \int_0^{\infty} d\bar{E} \int_{-\bar{E}}^{\bar{E}} d\omega \: \omega^{2n} e^{-\beta \bar{E} + S(\bar{E} + \omega/2)+S(\bar{E} - \omega/2)} |\hat{O}(\bar{E},\omega)|^2
\end{align}
where $O(\bar{E},\omega)$ is the matrix element $\hat{O}_{lm}$ expressed in $(\bar{E},\omega)$ coordinates. Approximating
\begin{align}
    S(\bar{E} + \omega/2)+S(\bar{E} - \omega/2) \approx 2S(\bar{E}) + \frac{\omega^2}{4}S''(\bar{E})
\end{align}
we get:
\begin{align}
    \mu_{2n} \sim  \int_0^{\infty} d\bar{E} \: e^{-\beta \bar{E}}\int_{-2\bar{E}}^{2\bar{E}} d\omega \: \omega^{2n} e^{2S(\bar{E}) + \omega^2 S''(\bar{E})/4} |\hat{O}(\bar{E},\omega)|^2
\end{align}
By the extensivity of the function of the entropy and energy functions, the inverse of the second derivative $S''(\bar{E})$ is extensive, thereby yielding:
\begin{align}
    S''(\bar{E})\: \omega^2 \approx 0
\end{align}
for $\omega<<1$. However, when $\omega \rightarrow \infty$, the structure function $f_O(\bar{E},\omega)$ needs to vanish for large $\omega$ (as detailed in the main text). Therefore, we get:
\begin{align}
    e^{S''(\bar{E})\omega^2/4} \sim O(1)
\end{align}
which yields:
\begin{align}
    \mu_{2n} \approx \frac{ \int_0^{\infty} d\bar{E} \: e^{S(\bar{E})-\beta \bar{E}}\int_{-2\bar{E}}^{2\bar{E}} d\omega \: \omega^{2n} |f_{O}(\bar{E},\omega)|^2}{\int_0^{\infty} d\bar{E} \: e^{S(\bar{E})-\beta \bar{E}}}
\end{align}
For an operator $\hat{O}$ in a random-matrix ensemble, the matrix elements are of the form 
\begin{align}
    O_{lk} = \frac{1}{\sqrt{\mathcal{D}}} R_{lk}
\end{align}
as in eq. (12) in the main text, where $R_{lk}\sim \mathcal{N}(0,1)$ and $\mathcal{D}$ is the Hilbert-space dimension. Averaging over the randomness yields the following:
\begin{align}
    \langle \mu_{2n}\rangle &= \beta\int_0^{\infty} d\bar{E} \:e^{-\beta \bar{E}} \int_{-2\bar{E}}^{2\bar{E}} d\omega \: \omega^{2n} \\
    &= 2 \int_{0}^{\infty}d\omega \: \omega^{2n} e^{-\beta|\omega|/2} \nonumber\\
    &= \frac{2^{2n+2}(2n)!}{\beta^{2n+1}}  \nonumber\\
    &\sim \left( \frac{4n}{\beta e}\right)^{2n}
\end{align}

When the matrix elements decay as $O_{lk} \sim e^{-\gamma |\omega|}$, we instead have:
\begin{align}
    \langle \mu_{2n}\rangle 
    &= 2 \int_{0}^{\infty}d\omega \: \omega^{2n} e^{-(\beta/2+\gamma)|\omega|} \nonumber\\
    &= \frac{2^{2n+2}(2n)!}{(\beta+4\gamma)^{2n+1}}  \nonumber\\
    &\sim \left( \frac{4n}{(\beta+4\gamma) e}\right)^{2n}
\end{align}

\bibliography{citations}
\bibliographystyle{unsrt}